\documentclass[a4paper]{jpconf}
\usepackage{graphicx}
\def\26{{$\sigma_{26}$}}

\newcommand{\be}{\begin{equation}}
\newcommand{\ee}{\end{equation}}
\newcommand{\bea}{\begin{eqnarray}}
\newcommand{\eea}{\end{eqnarray}}
\newcommand{\sigmav}{$\langle\sigma v\rangle$}

\newcommand{\gsim}{\mathrel{\mathop{\kern 0pt \rlap
  {\raise.2ex\hbox{$>$}}}
  \lower.9ex\hbox{\kern-.190em $\sim$}}}

\begin{document}
\title{Conservative upper limits on WIMP annihilation cross section from Fermi-LAT $\gamma$-rays}

\author{Fiorenza Donato}

\address{Dipartimento di Fisica Teorica,  Torino University
and INFN, Via P. Giuria 1, 10122 Torino, Italy }\ead{donato@to.infn.it}

\author{Francesca Calore}

\address{II. Institute for Theoretical Physics, University of Hamburg, Luruper Chaussee 149, 22761 Hamburg, Germany}


\author{Valentina De Romeri}

\address{ IFIC (CSIC - Universidad  de Valencia), C/ Catedratico 
Jose Beltran 2, E-46980 Paterna (Valencia), Spain}


\begin{abstract}
The spectrum of an isotropic extragalactic $\gamma$-ray background (EGB)  
has been measured by the Fermi-LAT telescope at high latitudes.
Two new models for the EGB are derived from the subtraction of
 unresolved point sources and extragalactic 
diffuse processes, which could explain from 30\% to 70\% of the Fermi-LAT 
EGB.  
Within the hypothesis that the two residual EGBs are entirely due to the  
annihilation of dark matter (DM) particles in the Galactic halo, we   
obtain $conservative$  upper limits on their annihilation cross section \sigmav. 
Severe bounds on a possible Sommerfeld enhancement of 
the  annihilation cross section are set as well. 
Finally, would {\sigmav} be inversely proportional to the WIMP velocity, 
very severe limits are derived for the velocity-independent part of the 
annihilation cross section.
\end{abstract}

\section{The extragalactic $\gamma$-ray background} 
\label{EGB} 
The high latitude  ($|b|>10^{\rm o}$)
$\gamma$-ray emission measured by Fermi-LAT \cite{LATIsotropicSpectrum},
given its reduced contamination by galactic sources, 
is a powerful tool to set limits on the contribution of DM to the measured 
flux. The spectrum has been obtained after the subtraction from the data of 
 the sources resolved by the telescope, the (indeed model dependent) 
 diffuse galactic emission, 
the cosmic ray (CR) background in the detector and the solar $\gamma$-ray emission. 
For each low--flux source there may be a large number of \textsl{unresolved} 
point sources which have not been detected because of selection
effects, 
or too low emission. 

Most of the unassociated high latitude sources are blazars,  a class of Active Galactic Nuclei
(AGNs), and they pile to the EGB with the largest flux \cite{2010ApJ...720..435A}.  Galactic
resolved pulsars and Milli-Second Pulsars (MSPs) represent the second largest population  in
the Fermi-LAT catalog \cite{1FGL,2010ApJS..187..460A} and they are expected to contribute
significantly to the putative EGB. A non-negligible $\gamma$-ray flux seems to be guaranteed
by unresolved normal star-forming galaxies \cite{EGBStarformgal2010}.  Ultra-high energy CRs
(UHECRs) may induce secondary electromagnetic cascades, originating neutrinos and
$\gamma$-rays at Fermi-LAT energies \cite{EGBCMB1}. 
Unresolved blazars and MSPs are believed to contribute at least few percent
to the Fermi-LAT  EGB, while predictions for star-forming galaxies and UHECRs are highly model
dependent. 
In the following, we describe few classes of $\gamma$-ray emitters whose 
unresolved flux is firmly estimated in a non-negligible Fermi-LAT EGB percentage. 
In a conservative scenario (Model I), we will subtract AGN and MSPs
 to the Fermi-LAT EGB as derived in Ref. \cite{LATIsotropicSpectrum}. 
A more relaxed model (Model II) will be drawn by the further subtraction 
of a minimal flux from star-forming galaxies and CRs at the highest 
energies. 
The derivation of each contribution is described in details in \cite{CDD}.

\subsection{Models for the EGB} 
\label{sec_residui}
\begin{figure}[h] 
\begin{minipage}{20pc}
\includegraphics[width=20pc]{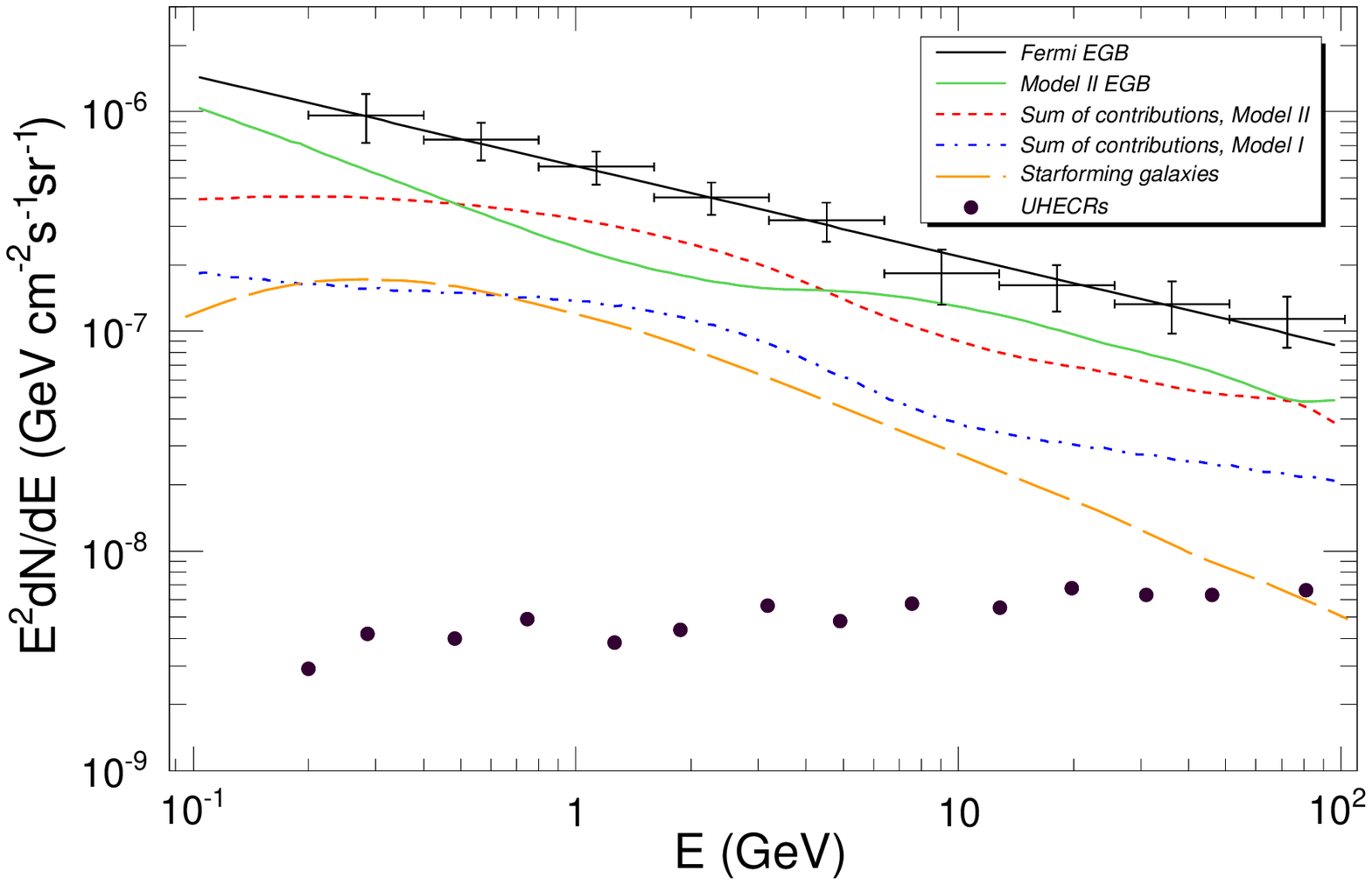}  
\caption{\label{fig:EGB_Model_II}$\gamma$-ray spectrum for $|b|>10^{\rm o}$ latitudes. 
Fermi-LAT data points 
are displayed along with their power--law fit (solid black curve) \cite{LATIsotropicSpectrum}. 
Dots and long dashed-curve (yellow) correspond to the UHECRs  and star-forming galaxies $\gamma$-ray
fluxes, respectively. Dot-dashed (blue) curve: sum of 
BL Lacs, FSRQs and MSPs fluxes. Short-dashed (red): sum of all the unresolved components.
Solid (green) curve is derived 
by subtracting all the contributions to the Fermi-LAT result (Model II). }   
\end{minipage}\hspace{2pc}%
\begin{minipage}{20pc}
\includegraphics[width=20pc]{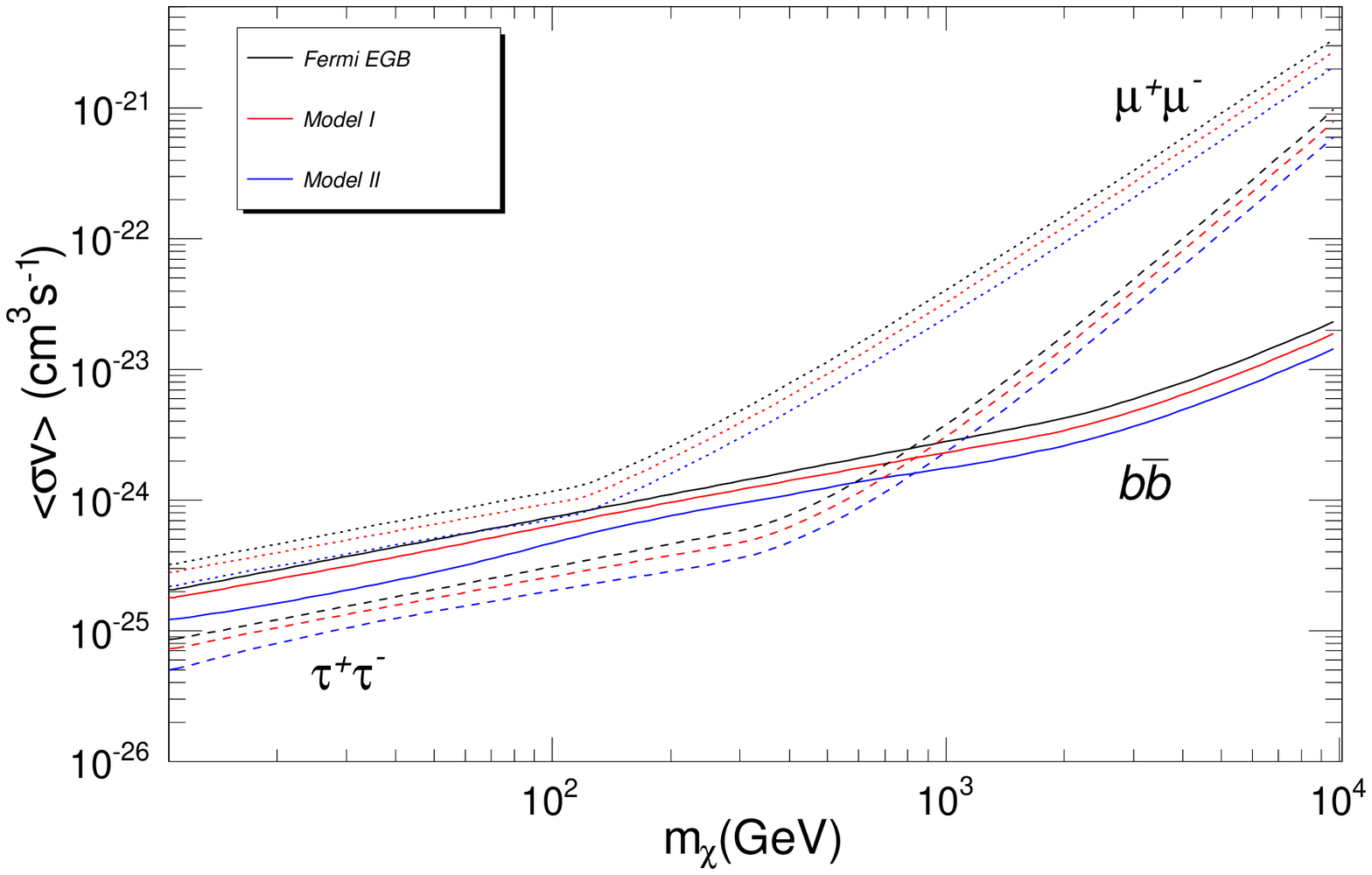}  
\caption{\label{fig:upper_bounds} Upper bounds
on {\sigmav} from $\gamma$-ray in the high latitude galactic halo, 
as a function of the DM mass. 
From top to bottom, solid lines refer to 90\% C.L.
 limits  from the comparison with  Fermi-LAT EGB   (black lines), 
Model I (red lines), Model II (blue lines) (see text for details). 
Dotted, solid and dashed lines correspond 
DM annihilation into $\mu^+\mu^-$, $b\bar{b}$, $\tau^+\tau^-$, respectively.}  
\end{minipage}
\end{figure}

The aim of this Section is to subtract
from the Fermi-LAT EGB \cite{LATIsotropicSpectrum} additional contributions 
from unresolved sources at latitudes $|b|>10^{\rm o}$.
The contributions removed  from the Fermi-LAT spectrum
are minimal. In fact, the predictions that we will take into account
for the relevant unresolved sources are the lowest 
ones according to the literature. 
For the Model I, we subtract from the Fermi-LAT EGB \cite{LATIsotropicSpectrum}
the unresolved contributions for both BL Lacs and FSRQs, 
and  MSPs. 
Model II for the EGB refers to the scenario where the 
additional contributions from star-forming galaxies and UHECRs
 add to explaining the Fermi-LAT EGB.  
For all the details, the interested reader is addressed to Ref. \cite{CDD}.
The ensuing fluxes are displayed in Fig. \ref{fig:EGB_Model_II}. 
At 100 MeV,  Model II explains about $70\%$ of the Fermi-LAT EGB,  
while above  1-2 GeV they count about $50\%$ of the total. 

\section{Upper bounds on DM annihilation cross section} 
\label{results} 
\label{sec:upperbounds}

We make the hypothesis that the residual fluxes in Fig. \ref{fig:EGB_Model_II}  
are entirely provided by the $\gamma$-rays produced by 
thermalized WIMP DM in the halo of the Milky Way. 
For the prediction of the DM flux we refer to \cite{CDD}, and
 simply remind that all the considered DM density profiles 
provide very similar results for latitudes well above the galactic plane. 
We derive upper bounds  at 90\% C.L. on the WIMP annihilation cross section 
from the   $\gamma$-ray Fermi-LAT EGB and the EGB residual 
fluxes identified as Model I and II. They are displayed in 
 Fig. \ref{fig:upper_bounds} and detailed in \cite{CDD}.
The subtraction of the minimal 
amount of $\gamma$-rays from unresolved sources lowers the 
limits on {\sigmav} by at least 50\%. 
 Our limits are {\it conservative}: it is very unlikely that a 
 higher {\sigmav} be compatible with Fermi-LAT EGB. Similarly, 
our upper limits could be lowered only
with  assumptions on non-homogeneous DM distributions or, of course, 
by comparing to a smaller EGB residual.

\subsection{Bounds on the Sommerfeld enhancement for {\sigmav}} 
\label{sect:Sommerfeld}
Recent claims on the excess of CR positrons \cite{2009Natur.458..607A} have  
stimulated the interpretation of data in terms of annihilating DM with fairly large 
annihilation cross sections of the order of $10^{-23}-10^{-22}$ cm$^3$/s. These numbers
are at least three orders of magnitude larger than the value indicated by observations
of the DM abundance due to thermal production. 
One way to boost the annihilation cross section is through the Sommerfeld effect 
\cite{somm,2009PhRvD..79a5014A,Lattanzi,hisano}, generically due to an attractive force acting between two particles,
$i.e.$ a Yukawa or a gauge interaction. 
In the case of DM particles, the main effect of such an attractive force would be to enhance {\sigmav}
by a factor proportional to $1/\beta = c/v$, where $v$ is the velocity of the DM particle
($1/v$ enhancement). The net result on the annihilation cross section writes as  
{\sigmav} = S {\sigmav}$_{0}$, where S sizes the Sommerfeld enhancement of the annihilation amplitude.
We have evaluated the Sommerfeld enhancement S
using the approximation of the Yukawa potential by the Hulthen potential, for which an analytic
solution is possible~\cite{2010JPhG...37j5009C,Feng2} 
(and checked that the solution coincides with the numerical one). 

\begin{figure}[h]
\begin{minipage}{19pc}
\includegraphics[width=19pc]{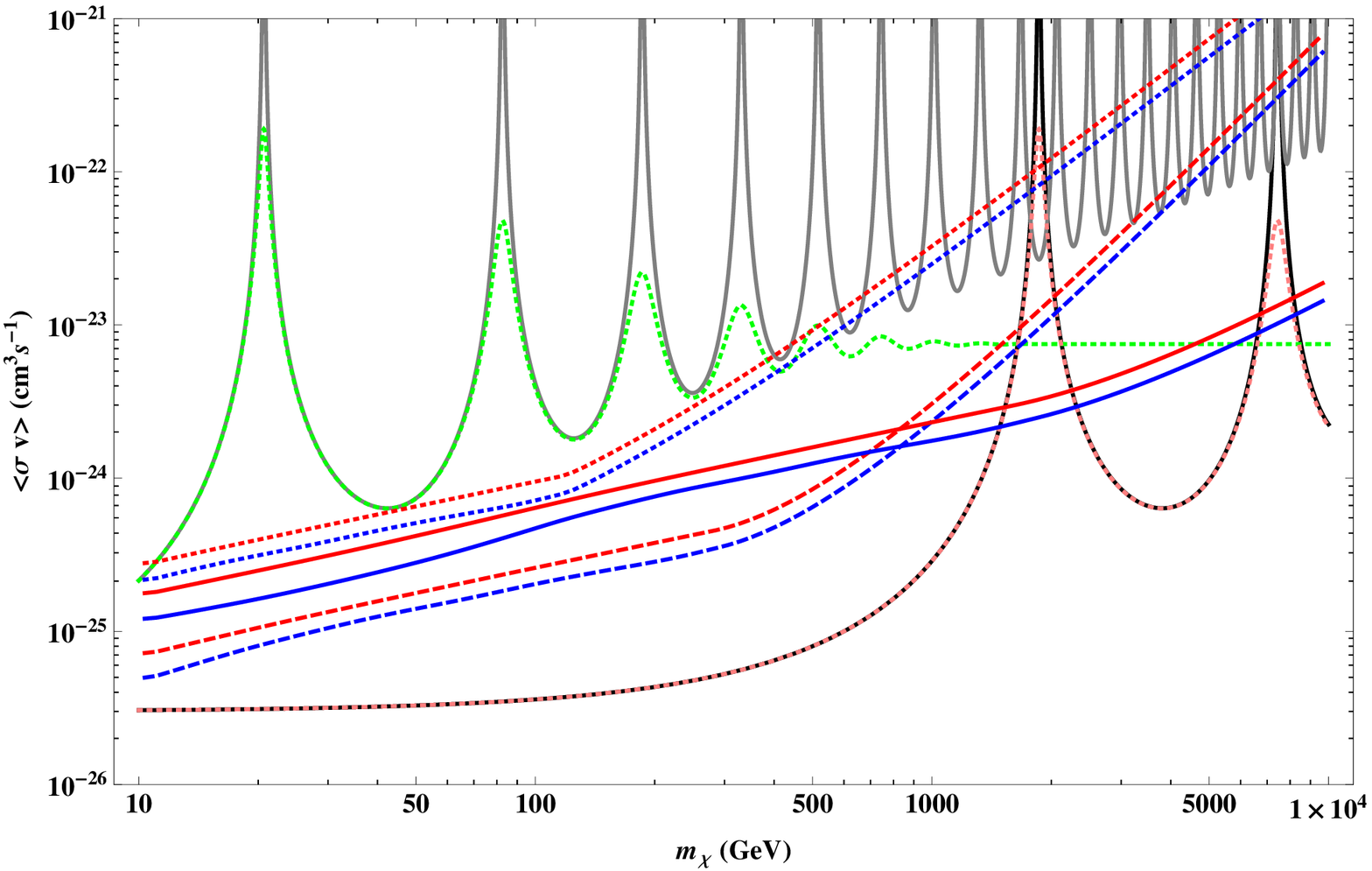}  
\caption{\label{fig:somm_sigmav} Sommerfeld enhanced {\sigmav} as a function of the DM mass, 
for $\alpha = \frac{1}{4\pi}$. Solid: $\beta =10 ^{-8}$, dotted: $\beta =10 ^{-3}$. 
The upper (lower) resonant curve is obtained for a force carrier of mass $m_{\phi} = 1$ GeV (90 GeV). 
The upper (lower) dotted, solid and dashed curves  correspond to the upper bounds for EGB Model I (Model II)
 in $\mu^+\mu^-$, $b\bar{b}$, $\tau^+\tau^-$, respectively (see Fig. \ref{fig:upper_bounds}.)} 
\end{minipage}\hspace{2pc}%
\begin{minipage}{19pc}
\includegraphics[width=19pc]{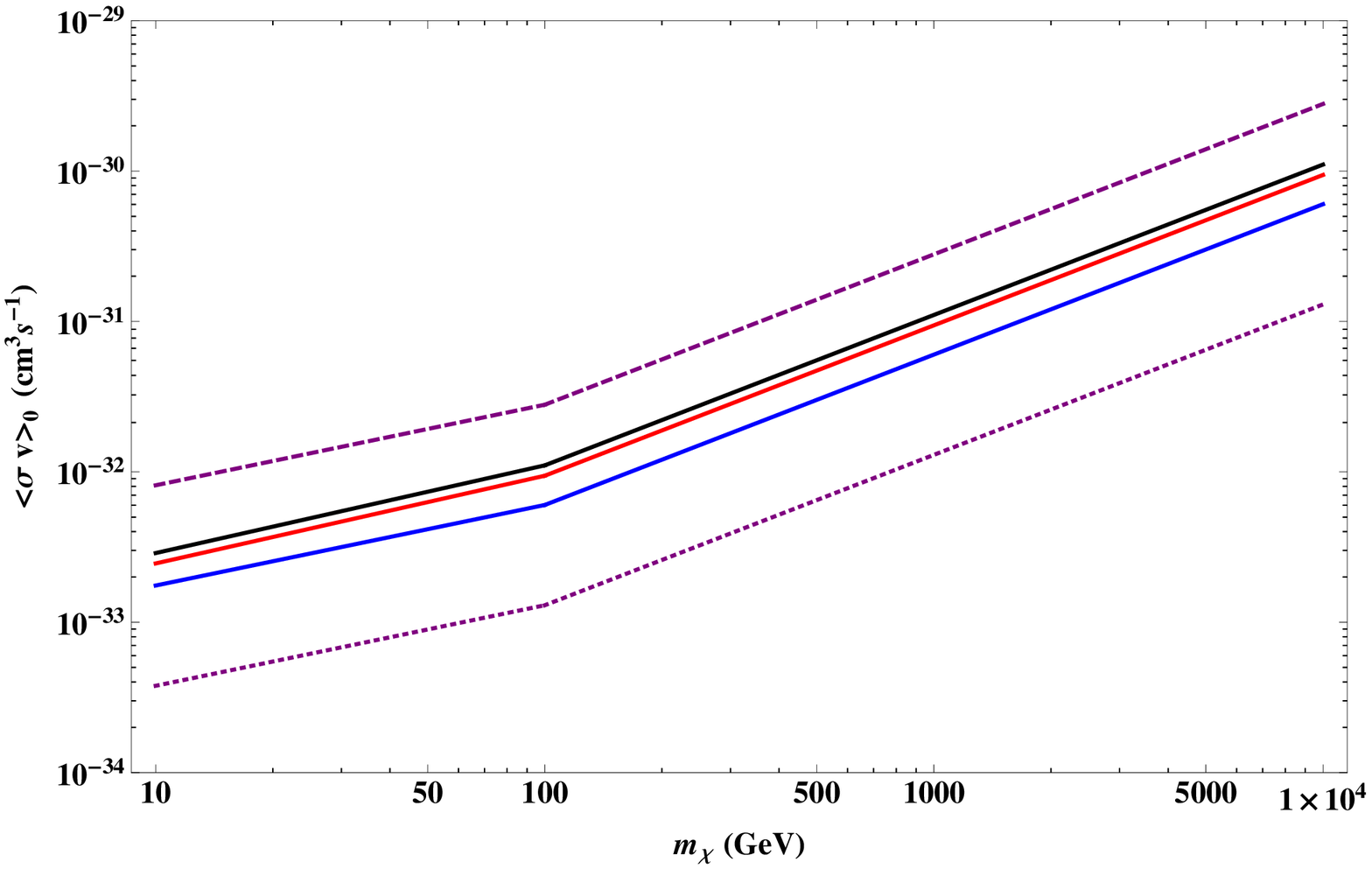}  
\caption{\label{fig:bounds_sigmav_0} Bounds on {\sigmav}$_{0}$ 
 as a function of the DM mass.
The central three bounds are obtained for $M_c = M_\oplus$, and from 
Fermi-LAT EGB (black line), Model I (red line) and 
Model II (blue line) respectively, from top to bottom.
The upper (lower) purple lines are derived for Model II EGB
and $M_c = 10^{2} M_\oplus (10^{-2} M_\oplus$).}  
\end{minipage}
\end{figure}

In Fig. \ref{fig:somm_sigmav}  we show the Sommerfeld 
enhanced cross sections with  over-imposed the upper bounds 
from the residual EGB Model I and Model II (see Fig. \ref{fig:upper_bounds}). Our results 
show that a Sommerfeld enhancement due to a force carrier of  $m_{\phi} < 1$ GeV 
(coupling $\alpha = \frac{1}{4\pi}$)  is strongly excluded by Model I and II for the Fermi-LAT EGB data. 
For a massive force carrier (90 GeV) only the resonant peaks above the TeV mass are excluded. 
The result holds for  $\beta =10 ^{-8}$  up to $\beta =10 ^{-3}$.

\subsection{Bounds from the high-redshift protohalos}
A possible way to boost the annihilation rate is to modify the 
particle theory and make the ansatz that the annihilation
cross section depends on the inverse of the velocity.
A boosted production of $\gamma$-rays in models with 
{\sigmav}$ \propto 1/v$ has been 
proposed for the first bound objects formed 
in the early phases of the universe \cite{proto,PK,CDD}. 
The velocity dispersion of the first protohalos that collapse at redshift $z_C$ 
is estimated to be very small ($\beta\sim 10^{-8}$) \cite{PK}. 
The photons arising from WIMP annihilations in very early 
halos can freely propagate with their energy red-shifting and 
reach the Earth in the range $\sim $ keV - TeV, while photons emitted out of this 
transparency window  are absorbed by the intergalactic medium. 
The $1/v$ enhancement of the annihilation cross section may be simply parameterized   \cite{PK}:
 $ \langle\sigma v\rangle=  \langle\sigma v\rangle_{0} \frac{ c}{v} \;\; {\rm cm^3/s}. $
We have evaluated the energy density in photons today from WIMP annihilation in the primordial halos 
and compared with the experimental photon density inferred for the 
Fermi-LAT, Model I and Model II EGB \cite{LATIsotropicSpectrum}, 
obtained by integrating the photon flux on the Fermi-LAT energy range (100 MeV - 100 GeV) \cite{CDD}. 
The results are displayed in Fig. \ref{fig:bounds_sigmav_0}. 
The bounds on {\sigmav}$_{0}$ are strong: for WIMP masses below 100 GeV it is forced to 
be $<10^{-33}$ cm$^3$/s. Upper bounds grow to $<10^{-32}$ cm$^3$/s  for $m_\chi\simeq$ 1 TeV
and sets to $<10^{-31}$ cm$^3$/s at 10 TeV. 
We make notice that they are more stringent than limits 
obtained from primordial light elements abundance and CMB anisotropies \cite{Hisano:2011dc}
and significantly improve the bounds of Ref. \cite{PK}.

 \section{Conclusions} 
The $\gamma$-ray EGB measured by Fermi-LAT \cite{LATIsotropicSpectrum} likely 
includes contributions from galactic and extragalactic $unresolved$ sources. 
We have discussed two residual EGB fluxes derived by the subtraction 
of unresolved BL Lacs, FSRQs and galactic MSPs, star-forming galaxies and UHECRs
to the $\gamma$-ray EGB measured by Fermi-LAT.
From our new residual EGB fluxes, we have set upper limits 
on the DM annihilation cross section into $\gamma$-rays. 
A conservative upper bound on {\sigmav} is derived by assuming 
that our new residual fluxes  are entirely due to WIMPs pair-annihilating 
in the halo of our Galaxy. 
Furthermore, we have shown these EGB residuals bound the Sommerfeld enhancement of {\sigmav}
 to a factor of 3-10-50-200 for
 $m_\chi$=10-100-1000-5000 GeV, respectively. In case of a Yukawa-like potential,
 a force carrier heavier than 1 GeV is required. 
Finally, within the hypothesis that {\sigmav} is inversely proportional to the WIMP velocity, 
very severe limits are derived for the velocity-independent part of the annihilation cross section.
\section*{References}
\bibliography{donato}

\end{document}